\documentclass[aps,prb,longbibliography,showpacs,twocolumn,superscriptaddress,amsmath,amssymb]{revtex4-2}
 
 \usepackage[utf8]{inputenc}
 \usepackage{amsmath,amsfonts,amssymb}
 \usepackage{pifont}
 \usepackage{rotating}
 \usepackage{graphicx}    
 \usepackage{epstopdf}
 \usepackage{color}
 \usepackage[caption=false]{subfig}
 \usepackage{soul}
 \usepackage{booktabs}
 \usepackage{hyperref}
 \definecolor{dark-red}{rgb}{0.9,0.15,0.15}
 \definecolor{dark-blue}{rgb}{0.15,0.15,0.4}
 \definecolor{medium-blue}{rgb}{0,0,0.5}
 \hypersetup{colorlinks=true,citecolor=blue,linkcolor=blue,urlcolor=blue}

\begin{document}
\title{Frustration-driven unconventional magnetism in the Mn$^{2+}$ ($S=\frac{5}{2}$) based two-dimensional triangular-lattice antiferromagnet Ba$_{3}$MnTa$_{2}$O$_{9}$}

\author{Romario Mondal}

\affiliation{Department of Physics, IIT (ISM) Dhanbad, Jharkhand 826004, India}

\author{Sk. Soyeb Ali}

\affiliation{Department of Physics, Bennett University, Greater Noida, Uttar Pradesh 201310, India}

\author{Saikat Nandi}

\affiliation{Department of Physics, IIT Bombay, Powai, Mumbai 400076, India}

\author{S. Chattopadhyay}

\affiliation{UGC-DAE Consortium for Scientific Research Mumbai Centre, 246-C CFB, BARC Campus, Mumbai 400085 India}

\author{S. Ga\ss}

\affiliation{Leibniz Institute for Solid State and Materials Research Dresden, Dresden D-01069, Germany}

\author{L. T. Corredor}

\altaffiliation{Present address: Faculty of Physics, Technical University of Dortmund, Otto-Hahn-Str. 4, D-44227 Dortmund, Germany, and Research Center Future Energy Materials and Systems (RC FEMS), Germany}

\affiliation{Leibniz Institute for Solid State and Materials Research Dresden, Dresden D-01069, Germany}

\author{A. U. B. Wolter}

\affiliation{Leibniz Institute for Solid State and Materials Research Dresden, Dresden D-01069, Germany}

\author{V. Kataev}

\affiliation{Leibniz Institute for Solid State and Materials Research Dresden, Dresden D-01069, Germany}

\author{B. B\"uchner}

\affiliation{Leibniz Institute for Solid State and Materials Research Dresden, Dresden D-01069, Germany}

\affiliation{Institute for Solid State and Materials Physics and W{\"u}rzburg-Dresden Cluster of Excellence ct.qmat, TU Dresden, D-01062 Dresden, Germany}

\author{A. Alfonsov}

\affiliation{Leibniz Institute for Solid State and Materials Research Dresden, Dresden D-01069, Germany}

\author{S. Wurmehl}

\affiliation{Leibniz Institute for Solid State and Materials Research Dresden, Dresden D-01069, Germany}

\author{A. V. Mahajan}

\affiliation{Department of Physics, IIT Bombay, Powai, Mumbai 400076, India}

\author{S. K. Panda}
\email[Email: ]{swarup.panda@bennett.edu.in}
\affiliation{Department of Physics, Bennett University, Greater Noida, Uttar Pradesh 201310, India}

\author{T. Dey}
\email[Email: ]{tushar@iitism.ac.in}

\affiliation{Department of Physics, IIT (ISM) Dhanbad, Jharkhand 826004, India}

 \begin{abstract}
 
 A triple perovskite oxide Ba$_{3}$MnTa$_{2}$O$_{9}$ has been synthesized and its magnetic properties have been investigated through dc and ac magnetization, specific heat, electron spin resonance (ESR) measurements, and density functional theory (DFT) calculations. Mn$^{2+}$ ($S$ = 5/2) ions are the only magnetic species present in the material. These Mn$^{2+}$ ions constitute a quasi-two-dimensional triangular network in the crystallographic $ab$-plane. Magnetization and specific heat measurements reveal the absence of any long-range magnetic order down to 0.5\,K despite the presence of antiferromagnetic correlations between the magnetic ions, suggesting the presence of geometric frustration in the material. The entropy release is lower than the expected theoretical value of $Rln(6)$, further suggesting the presence of frustration. First-principles calculations using density functional theory (DFT) and atomistic spin dynamics (ASD) simulations further support this lack of static magnetic order even at low temperatures and identify the competing magnetic interactions along with the quasi-2D magnetic dimensionality as the underlying origin of such an unconventional magnetic behavior.

\end{abstract}
				
\date{\today}

\maketitle

\section{Introduction}
 
   Over the last few decades, intense efforts have been made to understand the quantum- and thermal-fluctuation-induced non-trivial magnetic phenomena in various frustrated materials at low temperatures. These are likely to arise from the combined effect of geometric frustration, reduced dimensionality, spin degrees of freedom, and single-ion anisotropy (easy-axis and easy-plane type) \cite{ramirez, balents, Greedan2001}. One of the families of materials wherein interesting phase transitions occur and exotic magnetic ground states are stabilized is the quasi-two-dimensional (2D) triangular-lattice antiferromagnets (TLAF family) \cite{chubukov, collins}. The ground state of TLAFs at zero magnetic field is known to possess a non-collinear 120$^\circ$ spin configuration, which, with an increasing external magnetic field, evolves to the up-up-down (uud), canted, and finally a fully polarized state \cite{shirata, zhou}. The experimental realization of these states is well captured in the isothermal magnetization ($M$) versus magnetic field ($H$) curve in the form of plateaus. The most prominent one observed in several TLAFs is the plateau at 1/3 of the saturation magnetization \cite{zhou}.
   \par
   Recently, TLAFs with triple perovskite framework A$_{3}$MM$^\prime$$_{2}$O$_{9}$, where A = Ba, Sr, Ca; M = Co, Ni, Mn, and M$^\prime$ = Sb, Nb, Ta, were explored and proposed to comprise of triangular layers of magnetic M ions separated by layers of non-magnetic A and M$^\prime$ ions. Ba$_{3}$CoSb$_{2}$O$_{9}$ is the most celebrated $S$ = 1/2 TLAF where Co$^{2+}$ ions provide the experimental realization of theoretically predicted quantum-fluctuation-driven successive phase transitions and exotic spin states \cite{shirata, zhou, susuki, koutroulakis}. The isostructural sister compounds Ba$_{3}$CoNb$_{2}$O$_{9}$ \cite{yokata, lee} and Ba$_{3}$CoTa$_{2}$O$_{9}$ \cite{ranjith} have also exhibited similar low-temperature magnetic properties. In the $S$ = 1 TLAF,  Ba$_{3}$NiSb$_{2}$O$_{9}$, evidence of a double phase transition and magnetization plateaus are reported \cite{Shirata2}. However, the Nb-analog, Ba$_{3}$NiNb$_{2}$O$_{9}$, exhibits a single phase transition, which is ascribed to the prominent quantum fluctuations and easy-plane anisotropy \cite{Hwang, Lu1}. Some TLAFs with a high-spin value $S$ = 5/2 were also investigated recently, aiming to understand their magnetic phase transitions and how thermal fluctuations distinguish them from their quantum counterparts (with $S$ = 1/2 and 1). Ba$_{3}$MnSb$_{2}$O$_{9}$ \cite{Shu2023, Doi2004, SUN2015, TIAN2014} exhibits a single-phase transition from paramagnetic to a long-range antiferromagnetic state in which the $S$ = 5/2 Mn$^{2+}$ ions form a 120$^\circ$ spin structure in the crystallographic $\emph{ab}$-plane of the hexagonal crystal lattice, which is concluded to be the effect of easy-plane anisotropy. On the other hand, the Nb-analog, Ba$_{3}$MnNb$_{2}$O$_{9}$ \cite{lee2014, jiao} shows a two-step phase transition with weak easy-axis anisotropy.
   \par
   It would be interesting to substitute the M$^\prime$-site with a different cation and study the resulting magnetic and structural phase transitions. In this work, we have prepared and investigated the magnetic properties of a novel $S$ = 5/2 frustrated TLAF, Ba$_{3}$MnTa$_{2}$O$_{9}$ (later mentioned as BMTO), through X-ray diffraction (XRD), dc and ac magnetization, specific heat ($C_{p}$), electron spin resonance (ESR), density functional theory (DFT) calculations, and spin dynamics simulations. We note that J.W. Chen \emph{et} \emph{al.} \cite{chenBa3MnTa2O9} studied dielectric properties of BMTO but its crystal structure and magnetic properties are yet to be investigated. Our experimental and theoretical findings suggest that BMTO hosts a quasi-2D edge-shared triangular lattice of the magnetic Mn$^{2+}$ ($S$ = 5/2) ions. This triangular lattice is parallel to the crystallographic $ab$-plane. The nearest neighbor interaction is antiferromagnetic leading to magnetic frustration in this material. BMTO evades any glassy and long-range static magnetic ground state at least down to 500\,mK despite strong in-plane magnetic correlations. Our combined experimental and theoretical investigation strongly suggests that BMTO could host an exotic and unconventional magnetic state, driven by frustration and low dimensionality.
     
  \section {Experimental Details and Computational Methods}
Polycrystalline samples of BMTO and its nonmagnetic analog Ba$_3$ZnTa$_2$O$_9$  (later mentioned as BZTO) were synthesized by a conventional solid-state reaction method. Stoichiometric amounts of the precursors BaCO$_3$ (99.99\%), MnO (99.99\%), ZnO (99.99\%), and Ta$_2$O$_5$ (99.85\%) were thoroughly ground for $\sim$2 h using an agate mortar and pestle. The mixtures were calcined at 900$^{\circ}$C for 12\,h and then sintered at 1150$^{\circ}$C (for BMTO) and 1200$^{\circ}$C (for BZTO) for $\sim$3 days with several intermediate grinding and pelletization steps. After each sintering step, the furnace was slowly cooled down (120$^\circ$/h). 
\par
Powder X-ray diffraction (XRD) measurements were performed using a Rigaku Smartlab High-Resolution X-ray diffractometer (HR-XRD) with Cu-K$_{\alpha1}$  radiation ($\lambda$ = 1.5406\,\r A) at room temperature to check the phase purity of the sample. The Rietveld refinement of the structure model was carried out using the FULLPROF program \cite{RODRIGUEZ1993}. The refined parameters were put into a structural model and visualized in VESTA software \cite{Momma2008}.
\par
The temperature-dependent dc and ac susceptibility measurements were carried out in a Superconducting Quantum Interference Device-Vibrating Sample Magnetometer (SQUID-VSM, Quantum Design) in the temperature range of 1.8\,K $\leq$ $\emph{T}$ $\leq$ 300\,K and magnetic field range of 0 $\leq$ $H$ $\leq$ 70\,kOe. The magnetic susceptibility in the high-temperature (300\,K $\leq$ $\emph{T}$ $\leq$ 750\,K) and low-temperature (0.4\,K $\leq$ $\emph{T}$ $\leq$ 2\,K) ranges was measured using an oven and a $^3$He insert, respectively, in a Magnetic Property Measurement System (MPMS) from Quantum Design. A background subtraction was subsequently performed with the measured data. The isothermal magnetization at different temperatures in the magnetic field range of 0 $\leq$ $H$ $\leq$ 140\,kOe was measured using the VSM option in a Quantum Design Physical Property Measurement System (PPMS).
\par
Specific heat $C_p$ as a function of temperature (0.5\,K $\leq$ $\emph{T}$ $\leq$ 230\,K) was measured at various magnetic fields using a heat-pulse relaxation
technique in a Quantum Design PPMS. A $^3$He insert was used for measurements at $\emph{T}$ $<$ 1.8\,K. Note that the specific heat of the puck and the grease was subtracted from the total specific heat to obtain the specific heat of the sample.
\par 
ESR measurements were performed using a Bruker X-band ESR spectrometer with a fixed microwave frequency of $\nu$ = 9.56\,GHz. The magnetic field ($H$) was varied from 0 to 9\,kOe and, the ESR spectra (field derivatives of the microwave
power absorption) were recorded at several temperatures ranging from 4\,K to 300\,K. A $^4$He-gas flow cryostat (Oxford Instruments) was used for the control of the sample temperature.
\par 
To provide microscopic understanding of the observed magnetic behavior, we performed first-principles calculations using density functional theory (DFT)~\cite{DFT1,DFT2} with two complementary full-potential methods: the full-potential linearized augmented plane wave (FP-LAPW) method in WIEN2K code \cite{SCHWARZ2003259} and the full-potential linearized muffin-tin orbital (FP-LMTO) method in RSPt code~\cite{PhysRevB.12.3060,PhysRevB.91.125133,10.1007/3-540-46437-9_4}. Exchange-correlation effects were treated using the generalized gradient approximation (GGA) with an added Hubbard $U$ correction (GGA+$U$) for Mn-$d$ orbitals. Based on prior studies ~\cite{PhysRevB.102.094408, PhysRevB.86.165105}, $U = 4$\,eV and Hund’s $J = 0.8$\,eV were used. Brillouin-zone integration was performed on a $10\times10\times5$ $k$-mesh. 
\par
After achieving good charge-density convergence in our FP-LMTO calculations within the GGA+$U$ framework, the effective inter-site exchange parameters $J_{ij}$ between Mn spins were computed using the magnetic force theorem~\cite{magnetic-force1,magnetic-force2}. Here, the total energy is mapped onto a Heisenberg Hamiltonian of the following form: 
\begin{equation}
	\ H =  -\frac{1}{2}\sum_{i\neq j}J_{ij}(\hat{S}_i \cdot \hat{S}_j) 
	\label{H}
\end{equation}
where $\hat {S}_i$, $\hat {S}_j$ are spin unit vectors for the ${i}^{th}$ and ${j}^{th}$ site, respectively. The $J_{ij}$ values were extracted in a linear-response manner via a Green’s function formalism: 
\begin{equation}
	J_{i j}=\frac{\mathrm{T}}{4} \sum_n \operatorname{Tr}\left[\hat{\Delta}_i\left(i \omega_n\right) \hat{G}_{i j}^{\uparrow}\left(i \omega_n\right) \hat{\Delta}_j\left(i \omega_n\right) \hat{G}_{j i}^{\downarrow}\left(i \omega_n\right)\right],
\end{equation}
where $\Delta_i$ is the onsite spin splitting, the spin-dependent intersite Green’s function is $G_{ij}$, and $\omega_n$ are the $n^{th}$ Fermionic Matsubara frequencies. A detailed discussion of the implementation of the magnetic force theorem in RSPt is provided in Ref.~\onlinecite{PhysRevB.91.125133}. This method has been successfully used for many other transition metal compounds~\cite{PhysRevB.94.064427,PhysRevB.109.035125,PhysRevB.106.L180408}. The obtained exchange parameters were used to construct the spin Hamiltonian for this system. This was further analyzed to  study finite-temperature magnetization dynamics using atomistic spin dynamics (ASD) simulations within the Stochastic Landau-Lifshitz-Gilbert (SLLG) framework in UppASD package ~\cite{PhysRevB.54.1019,Skubic_2008}. We primarily investigated equilibrium spin configurations, and magnetic phase transitions.
\section {Results and Discussion}
\subsection{XRD and Crystal Structure}

\begin{figure}[ht]
	\begin{center}
		\includegraphics[width=\linewidth, trim={0.6cm 8.4cm 9.3cm 0.2cm,clip}]{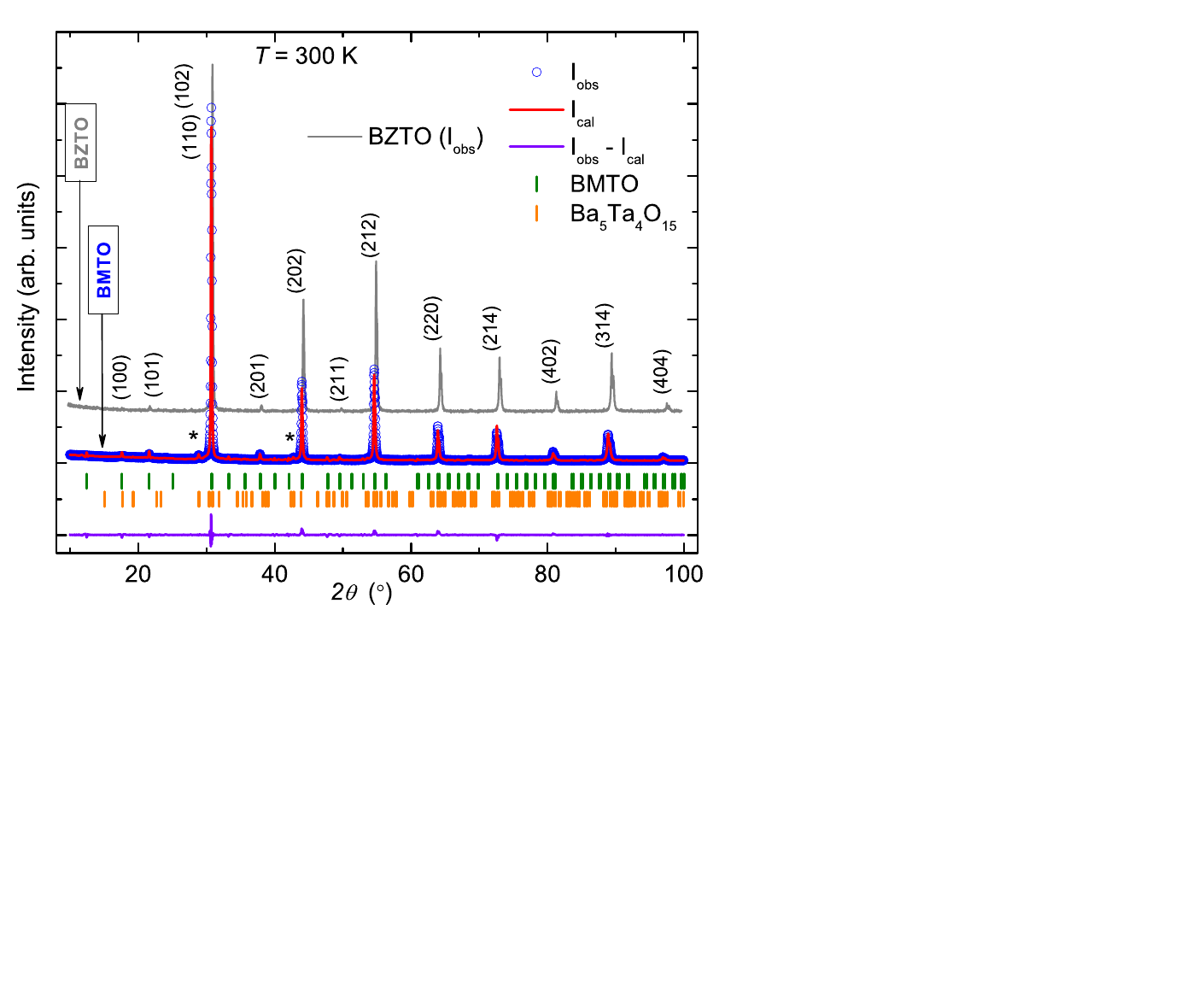}
		\caption{The room temperature powder x-ray diffraction data (blue circle) along with its double-phase Rietveld refinement (red solid line). Bragg positions for the main phase (BMTO) and the small fraction of non-magnetic secondary phase Ba$_5$Ta$_4$O$_{15}$ are shown as vertical lines in green and orange, respectively. Both phases belong to the same space group ($P\overline{3}m1$). Peaks corresponding to the Ba$_5$Ta$_4$O$_{15}$ phase are marked with ($\ast$). The violet solid line indicates the difference between the observed and calculated intensities. The  XRD pattern of the non-magnetic analog BZTO is shown with a solid gray line. This pattern is shifted along the $y$-axis for clarity.}
		\label{fig: x-ray diffraction}
	\end{center}
\end{figure}
Fig. \ref{fig: x-ray diffraction} shows the ambient temperature ($T$ = 300\,K) polycrystalline XRD pattern of BMTO (blue circles) and BZTO (grey line). Both patterns are very similar, suggesting that both compounds crystallize in the same structure. Treiber \textit{et al.} reported the formation of an ordered, trigonal Ba$_3$MnTa$_2$O$_9$ above 1200$^{\circ}$C \cite{TK82}, in which a specific ordering sequence of Mn and Ta (...Mn-Ta-Ta-Mn-Ta-Ta...) is constituted. As our synthesis conditions are also in line with them, BMTO and BZTO are expected to crystallize into the trigonal structure as well. Furthermore, since their sister compound, Ba$_3$MnNb$_2$O$_9$, has been reported to crystallize in an B-site ordered (...Mn-Nb-Nb-Mn-Nb-Nb...) trigonal crystal structure with space group $P\overline{3}m1$ \cite{xin2018, Liu2006}, and the fact that the ionic radii of Nb$^{5+}$ and Ta$^{5+}$ in an octahedral environment ($r$ = 0.64\,\r A \cite{Shannon1976}) are very close, a structure model similar to Ba$_3$MnNb$_2$O$_9$ was used as an educated guess to fit the room-temperature XRD pattern of Ba$_3$MnTa$_2$O$_9$ (for detailed analysis, see \cite{supp_material}). Following this approach, we were able to assign most of the reflections in the pattern of BMTO to the trigonal lattice by the Rietveld refinement with $P\overline{3}m1$ space group (green vertical lines in Fig. \ref{fig: x-ray diffraction}). However, two tiny additional reflections (marked with $\ast$ in Fig. \ref{fig: x-ray diffraction}) at (2$\theta$) 29$^\circ$ and 43$^\circ$ are found, which cannot be assigned to the BMTO crystal symmetry, suggesting that they account for a small fraction of a secondary phase. By careful phase matching with Rigaku smartlab software \cite{RigakuSmartLab}, these two additional reflections were identified as the strongest reflections of Ba$_5$Ta$_4$O$_{15}$, which orders in a trigonal crystal lattice as well. A refinement to the additional reflection lines (vertical orange lines) confirms that they indeed belong to the Ba$_5$Ta$_4$O$_{15}$ phase. From the relative intensities, it is estimated that the fraction of this additional phase is $\sim$1\% of the total sample volume. It should be noted that since this tiny fraction of secondary phase is non-magnetic, the magnetic properties of BMTO will not be affected by this phase. Therefore, our refinement (solid red line) strongly suggests the formation of a majorly single-phase BMTO with a small amount of non-magnetic additional phase. 
\par
Further, to check the degree of cation ordering between Mn$^{2+}$ and Ta$^{5+}$, their respective site occupancies were carefully varied during refinement maintaining the general stoichiometry of the triple-perovskite BMTO (as done in ref. \cite{TDey2012, W_Lee_2017, Tdey2013frustration}). As a result, we found that our BMTO sample may have a Mn-Ta anti-site mixing of $\sim$15\% (for the detailed analysis, see \cite{supp_material}). Hence, we conclude that the BMTO sample in this study orders in the trigonal structure ($P\overline{3}m1$) with a high degree of cation order. Owing to the large contrast in the neutron scattering lengths of Mn and Ta, a neutron diffraction experiment would be helpful to accurately estimate the anti-site mixing in BMTO.
\par
The room temperature XRD data of BZTO was also refined (not shown here) using the space group $P\overline{3}m1$. The refined structural parameters were obtained as $a$ = $b$ = 5.7959(7)\,\r A, $c$ = 7.0913(8)\,\r A, $\alpha$ = $\beta$ = 90$^\circ$, $\gamma$ = 120$^\circ$, in good agreement with a previous report \cite{BMN03}.
\par
\begin{figure}[ht]
	\begin{center}
		\includegraphics[width=\linewidth, trim={0cm 1.4cm 0cm 0cm,clip}]{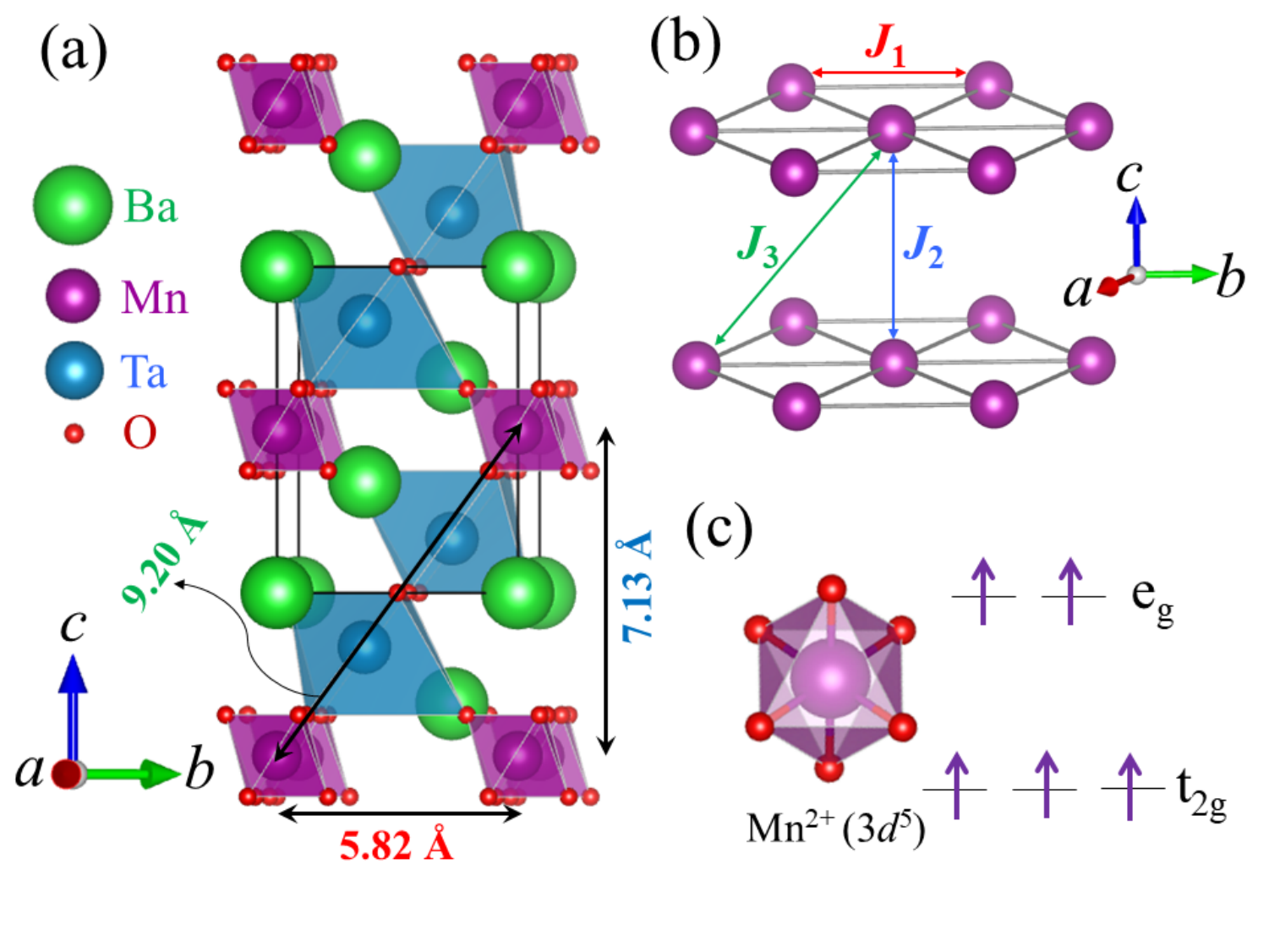}
		\caption{(a) The idealized crystal structure of BMTO is shown. The solid (black) lines indicate the unit cell. Mn$^{2+}$ ions (only at $1b$ site) and Ta$^{5+}$ ions (only at $2d$ site) are shown with the purple and light blue octahedral oxygen environment, respectively. Three thick double-sided arrows (horizontal, vertical, and slanted) indicate the distance between the first, second, and third nearest neighbor Mn$^{2+}$ ions, respectively. (b) Two edge-shared triangular layers of the Mn$^{2+}$ ions are shown with the exchange interaction between the first ($J_1$), second ($J_2$), and third ($J_3$) nearest neighbor Mn$^{2+}$ ions. (c) MnO$_{6}$ octahedra and the high-spin electronic configuration (3$d^5$) of Mn$^{2+}$ ions are shown.}
		\label{fig: structure}
	\end{center}
\end{figure}

The unit cell parameters and atomic positions obtained from the crystal structure refinement of BMTO data are listed in Table \ref{tab:str_parameter}. The structure of BMTO based on the refined structural model (without considering Mn-Ta anti-site mixing) is shown in Fig. \ref{fig: structure}(a), where the Mn$^{2+}$ ions (purple spheres) are the only magnetic ions in this material. Under the influence of an octahedral crystal field, they possess a high-spin state with $S$ = 5/2 (see Fig. \ref{fig: structure}(c)). In the fully ordered trigonal structure, Mn$^{2+}$ ions occupy the Wyckoff site 1$b$ (0, 0, 1/2) thereby forming an edge-shared triangular lattice in the crystallographic $ab$-plane, as shown in Fig. \ref{fig: structure}(b). These layers are stacked along the crystallographic $c-$axis separated by non-magnetic Ba$^{2+}$ ions and two layers of TaO$_6$ octahedra. Mn$^{2+}$  ions interact through superexchange interactions mediated by intermediate oxygen and tantalum orbitals \cite{jiao}. The first ($J_1$), second ($J_2$), and third ($J_3$) nearest neighbor interactions between the Mn$^{2+}$ ions are sketched in Fig. \ref{fig: structure}(b). The corresponding distances between the first, second, and third nearest Mn$^{2+}$ neighbors are $\sim$5.82\,\r A, $\sim$7.13\,\r A, and $\sim$9.20\,\r A, respectively. A comparatively larger ($\sim$7.13\,\r A) interlayer separation between the magnetic planes suggests a magnetically quasi 2D nature, while the edge-shared triangular lattice in the $ab-$plane (as shown in Fig. \ref{fig: structure}(b)) is expected to induce geometric frustration in this material.
\par
     
\begin{table}[h]
	\centering
	\caption{Rietveld refinement of the XRD data at $\emph{T}$ = 300\,K yields the structural parameters of BMTO as follows: crystal system: trigonal, space group: $P\overline{3}m1$ (No. 164), cell parameters: $a$ = $b$ = 5.8228(3)\,\r A, $c$ = 7.1252(4)\,\r A, $\alpha$ = $\beta$ = 90$^\circ$, $\gamma$ = 120$^\circ$. Atomic positions and their respective site occupancies in the unit cell are tabulated below. The goodness of fitting parameters are also provided.}
	\label{tab:str_parameter}
	\begin{ruledtabular}
		\begin{tabular}{ccccc}
			Atoms & $x/a$ & $y/b$ & $z/c$ & Occupancy \\
			(Wyckoff position) &  &  &  & \\
			\hline
			Ba1 ($1a$) & 0 & 0 & 0 & 0.0833 \\
			Ba2 ($2d$) & 1/3 & 2/3 & 0.6595(4) & 0.1667\\
			Mn ($1b$) & 0 & 0 & 1/2 & 0.0708\\
			Ta ($1b$) & 0 & 0 & 1/2 & 0.0125\\
			Ta ($2d$) & 1/3 & 2/3 & 0.1669(1) & 0.1542\\
			Mn ($2d$) & 1/3 & 2/3 & 0.1669(1) & 0.0125\\
			O1 ($3e$) & 1/2 & 0 & 0 & 0.2500\\
			O2 ($6i$) & 0.1100(8) & 0.2200(16) & 0.3736(9) & 0.5000\\
		\end{tabular}
	\end{ruledtabular}\\
	\vspace{1mm} 
	Bragg R-factor = 6.0\%, RF-factor = 6.7\%, R$_p$ = 14.2\%, $R_{wp}$ = 13.1\%, R$_{exp}$ = 4.6\%, $\chi^2$ = 8.1\\
	\vspace{1mm}
\end{table}
\subsection{Magnetization}

\begin{figure*}[t]
	\begin{center}
		\includegraphics[width=\linewidth, trim={1.4cm 8.2cm 10cm 0.4cm,clip}]{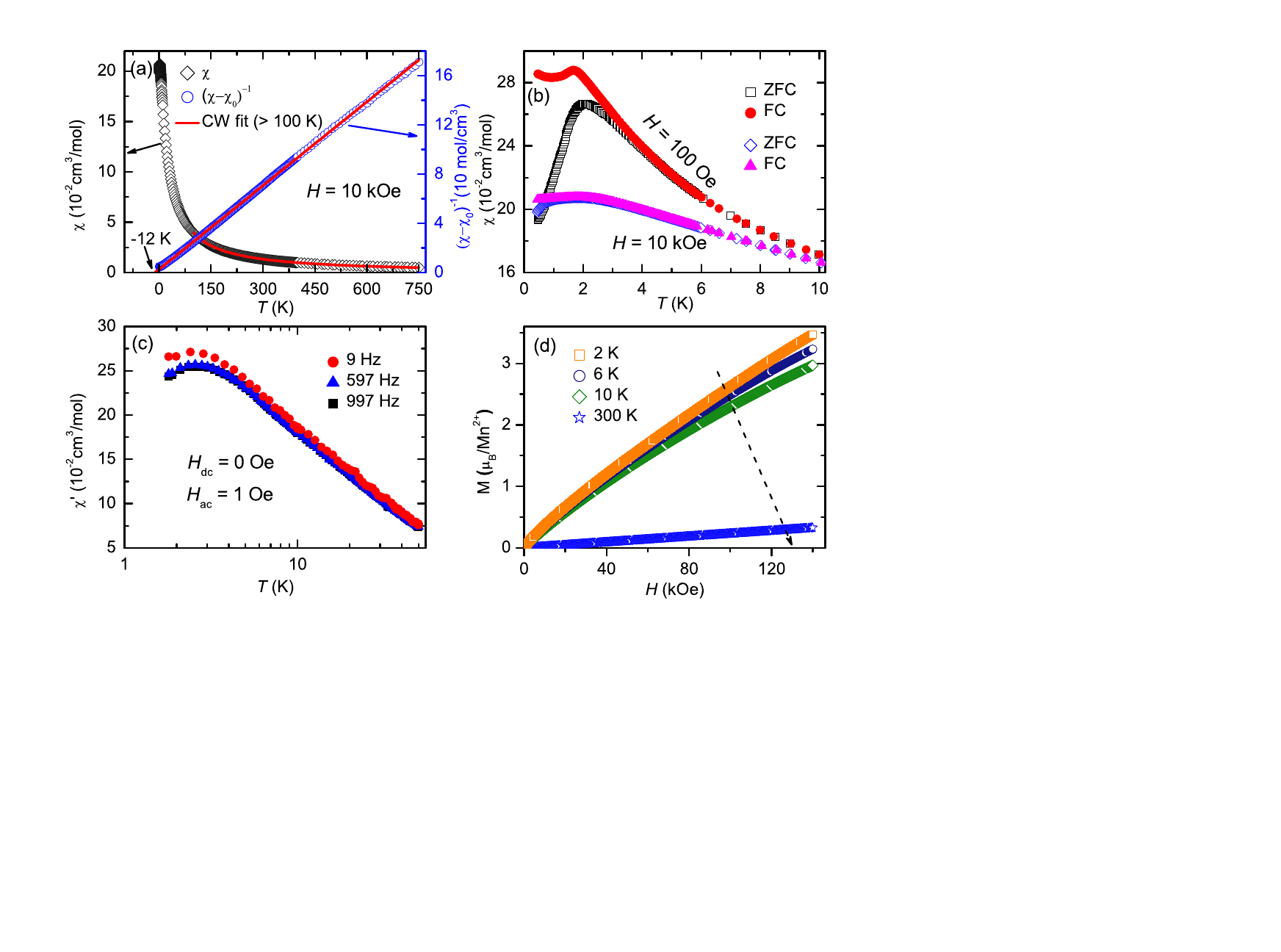}
		\caption{(a) Zero-field-cooled (ZFC) magnetic susceptibility measured with $H$ = 10\,kOe is shown as a function of temperature on the left $y$-axis. Its fitting with the Curie-Weiss (CW) law in the temperature range of 100\,K - 750\,K is shown by the solid red line. The inverse susceptibility $(\chi-\chi_0)^{-1}$ along with CW fitting (and its extrapolation) is plotted on the right axis. (b) The low-temperature part of the ZFC and field-cooled (FC) susceptibility measured with 100\,Oe and 10\,kOe are shown. (c) The ac-susceptibility ($\chi^{'}$) as a function of temperature measured with different frequencies is shown on a semi-log scale. (d) Isothermal magnetization curves at various temperatures measured with PPMS VSM (symbols) are shown. The arrow direction indicates increasing temperature.}
		\label{fig: magnetization}
	\end{center}
\end{figure*}

To understand the magnetic properties of BMTO, bulk magnetic measurements were carried out. The dc magnetic susceptibility data ($\chi$ = $\frac{M}{H}$) measured with $H$ = 10\,kOe are shown on left $y$-axis of Fig. \ref{fig: magnetization}(a) as a function of temperature ($T$) in the range 0.4\,K $\leq$ $\emph{T}$ $\leq$ 750\,K. A fitting of the high temperature (100\,K $\leq$ $\emph{T}$ $\leq$ 750\,K) susceptibility data with the Curie-Weiss (CW) law: 
\begin{equation}
	\label{eq:CWlaw}
	\chi(T) = \chi_0 + \frac{C}{(T - \theta_{CW})},
\end{equation}
is shown as the red solid line. Here, $\chi_0$ is the temperature-independent susceptibility and $\theta_{CW}$ is the Weiss temperature. The Curie constant, $C$ = $\frac{N_A\mu_{eff}^2}{3k_B}$, where $N_A$, $\mu_{eff}$, and $k_B$ are the Avogadro number, effective magnetic moment of free Mn$^{2+}$ ions, and Boltzmann constant, respectively. The parameters obtained from the fitting are C $\approx$ 4.4 cm$^{3}$-K/mol, $\chi_0$ $\approx$  -7.4$\times$$10^{-4}$ cm$^{3}$/mol, and $\theta_{CW}$ $\approx$ -12\,K. The effective magnetic moment $\mu_{eff}$ = $\sqrt{\frac{3k_BC} {N_A\mu_B^2}}$ $\mu_B$ = $\sqrt{8C}$\,$\mu_B$ $\approx$ 5.93\,$\mu_B$, is in close agreement with the expected spin-only value of $\mu_{eff}$ = $g\sqrt{S(S+1)}$\,$\mu_B$ $\approx$ 5.86\,$\mu_B$, where the Landé g-factor $g$ = 1.98 is obtained from our ESR measurement (discussed later) and $S$ (= 5/2) is the spin value of Mn$^{2+}$ ions in the high-spin state. The negative sign of $\theta_{CW}$ is an indication of a dominant antiferromagnetic interaction between the $S$ = 5/2 spins. Our DFT+U calculations also suggest a dominant intralayer first nearest neighbor ($J_1$) antiferromagnetic interaction (see Table \ref{tab: exchange}). 

The inverse susceptibility after subtracting the temperature-independent part $(\chi-\chi_0)^{-1}$ along with the CW fitting (solid straight line) are plotted on the right $y$-axis of Fig. \ref{fig: magnetization}(a).
\par
The low-temperature susceptibility measured at two different fields both in the zero-field-cooled (ZFC) and field-cooled (FC) mode is shown in Fig. \ref{fig: magnetization}(b). When measured with 10\,kOe field, the susceptibility saturates below $\sim$2.5\,K and a small ZFC-FC bifurcation is seen below $\sim$1\,K. At $H$ = 100\,Oe, a more prominent ZFC-FC bifurcation is observed below $\sim$4\,K with a broad maximum in ZFC data. A pronounced bifurcation in ZFC and FC susceptibilities could hint at long-range canted AFM order. However, no hysteresis behavior is observed at 2\,K. In the specific heat data (discussed later) also a broad maximum is found at $\sim$2.2\,K rather than a sharp peak. This suggests the absence of long-range order and the development of short-range spin correlations in the material. 
\par 
Disorder in a frustrated spin system can largely affect the ground state properties \cite{Dronova2022}. The Mn-Ta anti-site mixing could result in a spin-glass state \cite{Katukuri2015, Chakravarty2023}. The observed ZFC-FC bifurcation could be an indication of a spin-glass type behavior. In this case, one would expect a frequency dependency of the ac susceptibility. However, our ac susceptibility (in-phase component) measurement as a function of temperature $\chi^{'}(\emph{T})$ at three different frequencies (Fig. \ref{fig: magnetization}(c)), shows that the broad maximum at $\sim$2\,K is frequency independent. This indicates the absence of any glassy behavior in BMTO. The other possibility is that the characteristic frequency of a spin glass state is well below the lowest frequency measured (9\,Hz).
\par
The isothermal magnetization $M(H)$ up to $H$ = 140\,kOe at various temperatures is shown in Fig. \ref{fig: magnetization}(d). No hysteresis was observed, even at the lowest temperature. The $M(H)$ curve at $T$ = 300\,K is linear, indicating the paramagnetic nature of the material. At lower temperatures, a small curvature develops, but no indication of a sudden slope change is found. At 2\,K, the magnetization does not reach saturation even at 140\,kOe magnetic field. The magnetization at 140\,kOe is only about 71\% of the expected saturation value of 4.95\,$\mu_{B}$/Mn$^{2+}$ ($M_s = gS\mu_{B}$, where $g$ = 1.98 taken from the ESR analysis). Unlike some other TLAFs \cite{lee2014, shirata, Hwang}, no plateau-like behavior indicative of a field-induced spin state ($uud$) is observed in BMTO at least up to 140\,kOe. 

\subsection{Specific Heat}

\begin{figure*}[t]
	\begin{center}
	\includegraphics[width=\linewidth, trim={1cm 7.8cm 10cm 1cm,clip}]{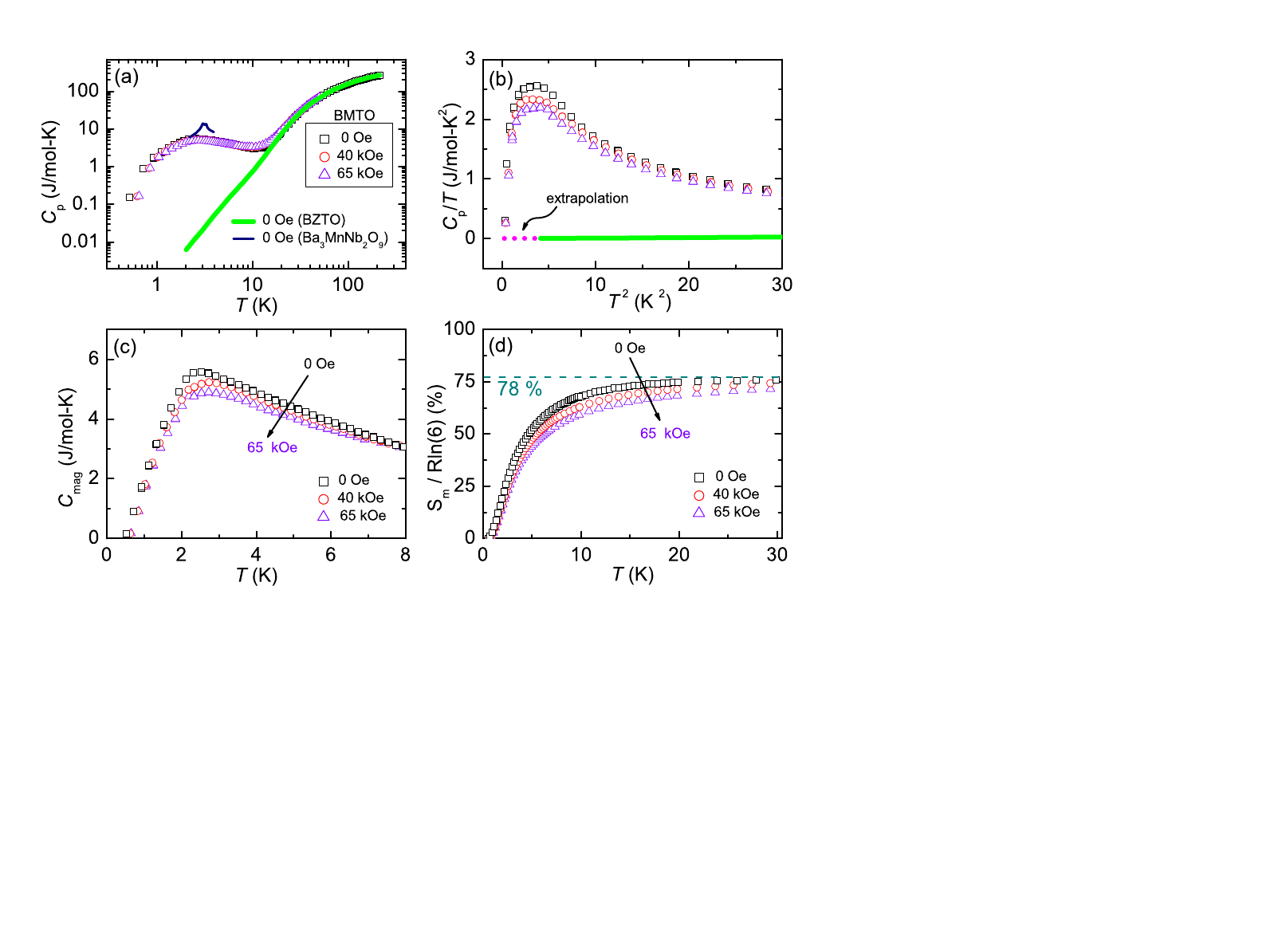}
	\caption{(a) Specific heat $C_{p}$ of BMTO and non-magnetic Ba$_{3}$ZnTa$_{2}$O$_{9}$ (BZTO) are shown as a function of temperature $(T)$ in log-log scale. Data at different applied fields for BMTO are shown by symbols and zero field data for BZTO is shown by the green solid line. The blue solid line indicates the zero field $C_{p}$ data of Ba$_{3}$MnNb$_{2}$O$_{9}$ adopted from Ref. \cite{lee2014} (b) The $C_{p}/T$ is plotted as a function of $T^2$ in the low-temperature region both for BMTO and BZTO. The straight line corresponding to BZTO is extrapolated in the range 0.5\,K - 1.8\,K (see text) and shown as a dotted line. (c) $C_{mag}$ of BMTO for different fields are plotted as a function of temperature. The solid arrow indicates the increasing field (d) The magnetic entropy for various applied fields is shown with an arrow that guides the rising of the fields. The horizontal dashed line indicates the maximum entropy released.}
	\label{fig: heat capacity}
	\end{center}
\end{figure*}

To get further insight into the ground state of BMTO, the specific heat $C_{p}(T)$ was measured at zero and finite magnetic fields in the temperature range 0.5\,K to 230\,K. The resulting data is plotted (symbols) in Fig. \ref{fig: heat capacity}(a) in a log-log scale. The specific heat of BZTO is measured at 0\,Oe in the temperature range 1.8\,K to 230\,K as a non-magnetic reference. As can be seen in Fig. \ref{fig: heat capacity}(a), the raw specific heat data of BZTO (solid green line) nicely match that of BMTO down to $\sim$30\,K. Below 30\,K, the BMTO data deviate, take an upturn, and produce a broad anomaly at $\sim$2\,K. A transition to a long-range ordered state can be ruled out since a sharp $\lambda$-like peak is absent. This is in contrast with its isostructural sister compound Ba$_{3}$MnNb$_{2}$O$_{9}$ which hosts a long-range ordered ground state \cite{lee2014, ZhaomingTian2014}. The $C_{p}(T)$ data of Ba$_{3}$MnNb$_{2}$O$_{9}$ (taken from Ref. \cite{lee2014}) is also shown in the same graph for a comparison. The low-temperature data for BMTO is shown as $C_{p}/T$ vs $T^2$ in Fig. \ref{fig: heat capacity}(b) where the broad anomaly is prominently visible. At higher magnetic fields, the height of the anomaly reduces but its position remains almost unchanged. Since BMTO is an insulating material as found in the calculated total density of states (discussed later), the total specific heat is expected to have a contribution from the magnetic and lattice parts only, i.e. $C_{p}$ = $C_{mag}$ + $C_{lat}$. To extract the magnetic specific heat ($C_{mag}$), one needs to subtract the lattice part from the total specific heat $C_{p}$. The specific heat data of non-magnetic BZTO can be considered as the lattice part ($C_{lat}$). At temperature below about 15\,K, $C_{p}$ data of BZTO follows $C_{p}$ = $\beta T^{3}$ and becomes a straight line in the $C_{p}/T$ vs $T^2$ plot as expected. We estimated the value of beta ($\beta = 0.747\,mJ\,mol^{-1}K^{-4}$) by fitting the BZTO data with $C_{p}$ = $\beta T^{3}$ in the range 2–10\,K. The Debye temperature was estimated to be $\theta_{D}$ = $(\frac{12\pi^4nR}{5\beta})^\frac{1}{3}$ $\approx$ 339\,K \cite{AWAKA2012471} (here $R$ is the molar gas constant and $n$ = 15 is the number of atoms in a formula unit of BMTO). The fitted curve was extrapolated to low temperatures (0.5\,K to 2\,K) and is shown as the dotted line in Fig.\ref{fig: heat capacity}(b). The combined curve in the temperature range 0.5\,K to 230\,K was taken as the lattice part ($C_{lat}$). To account for the mismatch in the molar masses of BMTO and BZTO, a scaling factor ($b$ = 0.98) was used \cite{Bouvier1991} to adjust the lattice part ($C_{lat}$). The obtained $C_{mag}$ at different fields is shown as a function of temperature in Fig. \ref{fig: heat capacity}(c). The broad anomaly at $\sim$2.2\,K appears to be nearly field-independent. It is worth noting that our computed temperature-dependent magnetic specific heat $C_{mag}(T)$ also shows a broad peak as displayed in Fig. \ref{fig: CmagDFT}(a). The broad peak presumably indicates the presence of short-range spin correlations and is in line with our magnetization results. Such behavior is also found in other frustrated magnetic materials \cite{TDey_Ba3InIr2O9_PRB, Li_YbMgGaO4_Scireports}. 
\par
The Mn-Ta anti-site mixing could give rise to orphan spins as found in other frustrated systems \cite{TDey2012}. In this case, one would expect a shift in this broad anomaly (Schottky effect) with applied field which is not seen in our measurement. This suggests the absence of such orphan spins. Also, the absence of a Curie-tail in susceptibility at low temperature further confirms the absence of orphan spins (see Fig. \ref{fig: magnetization}(b)). Hence, we conclude that even though Mn-Ta anti-site mixing is present, our measurements are not affected by orphan spins.

\par 
The magnetic entropy $S_{m}$ (= \(\int_{0}^{T} \frac{C_{mag}}{T'} \,dT'\)) was estimated for different fields by integrating $C_{mag}/T$ with respect to temperature up to 30\,K. $S_m$/$Rln(6)$ (here, $Rln(6) = 14.9$\,J/mol-K) for different fields are plotted as a function of temperature in Fig. \ref{fig: heat capacity}(d). The magnetic entropy does not attain the theoretically expected value of $Rln(6)$ even up to 30\,K, which is well above the anomaly observed. The recovered entropy is only $\sim$78\%, suggesting that the entropy is released over an extended range of temperature as seen in other frustrated magnetic materials \cite{Bhattacharya_Sr3CuTa2O9_2024, TDey_Ba3InIr2O9_PRB}. 

\subsection{Electron Spin Resonance (ESR)}

\begin{figure}
\begin{center}
	\includegraphics[width=\linewidth, trim={0.6cm 8.5cm 5.5cm 0.6cm,clip}]{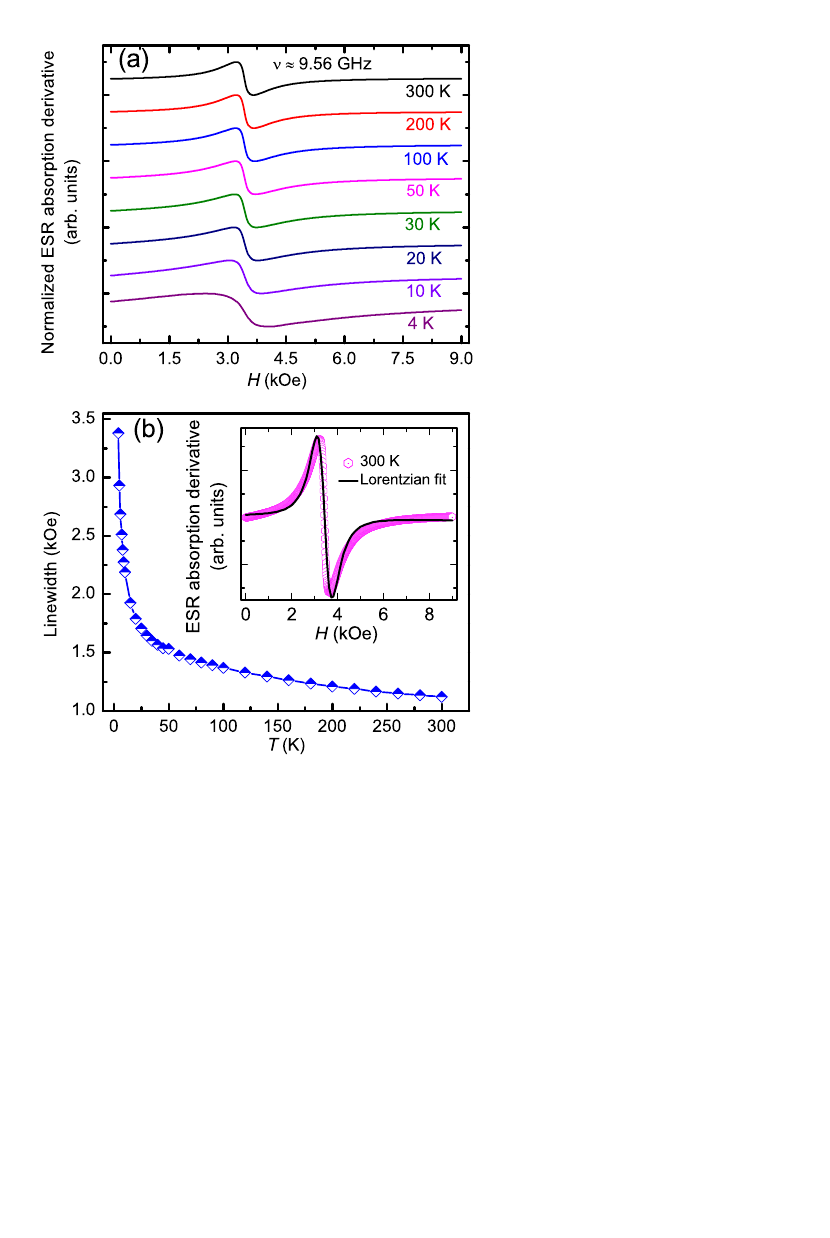}
	\caption{(a) The normalized ESR spectra at different temperatures are presented. (b) The linewidth as a function of temperature is plotted. Inset shows the Lorentzian fit to the ESR signal at $T$ = 300 K.}
	\label{fig: ESR}
	\end{center}
\end{figure}

Fig. \ref{fig: ESR}(a) shows the normalized ESR spectra in the field range of 0-9\,kOe recorded at several temperatures (4-300\,K) with a fixed microwave frequency of 9.56 GHz. Typically, the variation of the ESR linewidth with temperature carries information about the spin correlations in the sample. The ESR spectra recorded at different temperatures were fitted with a Lorentzian function to estimate the resonance field, linewidth, and g-factor. The inset of Fig. \ref{fig: ESR}(b) shows the fit of the $T$ = 300\,K data, which yields the $g$-factor to be 1.98 ($g$ = $\frac{h\nu}{H_{res}\mu_{B}}$, where $H_{res}$ and $h$ are the resonance field, and the Planck's constant, respectively). This value is close to the expected one for the exchange-coupled spin-only 3d Mn$^{2+}$ ions in an insulating host \cite{abragam2012electron}. Fig. \ref{fig: ESR}(b) represents a systematic increase in linewidth obtained from the fitting as the temperature decreases to the lowest temperature (4\,K). The linewidth in the range of 50-300\,K grows slowly with decreasing temperature, and below $\sim$50\,K, it starts to increase rapidly. The rapid increase signifies that the correlation between the Mn spins becomes stronger at low temperatures, as evidenced in our thermodynamic measurement (specific heat) discussed earlier. 

\subsection{Theoretical Results}

Our experimental results suggest the absence of any long-range magnetic order in BMTO down to 0.5 K. Nevertheless, the presence of short-range magnetic correlations is evident in the low-temperature regime. To understand the origin of such an unconventional behavior in the family of magnetic triple perovskite systems with classical spin, especially in light of frustration and low dimensionality, we performed detailed calculations using GGA+U approach. Our calculations were performed using the fully ordered trigonal structure of BMTO, consistent with the experimentally observed high degree of cation order in the sample. We note here that this is a computationally tractable and physically meaningful model to capture the essential electronic and magnetic properties of the system. We computed the total energies of two distinct spin configurations using a $2 \times 2 \times 1$ supercell. In the ferromagnetic (FM) state, Mn spins in the $ab$-plane are aligned parallel, whereas in the antiferromagnetic (AFM) configuration, Mn spins are antiferromagnetically aligned along the $a$- and $b$-directions. Our GGA+$U$ calculations reveal that the AFM state is energetically favored by 18 meV per formula unit, indicating the dominance of antiferromagnetic exchange interactions. The computed Mn spin moment is 4.5 $\mu_B$, consistent with a high-spin Mn$^{2+}$ ($S=5/2$) state, highlighting the localized nature of Mn-$d$ orbitals in agreement with the static magnetization and ESR data. Such strong localization is a characteristic feature observed in other Mn-based oxide systems, such as Ba$_3$MnSb$_2$O$_9$~\cite{Shu2023}, CaMn$_2$P$_2$ \cite{mallick2025neel} etc.

\begin{figure}
		\begin{center}
	\includegraphics[width=1\columnwidth]{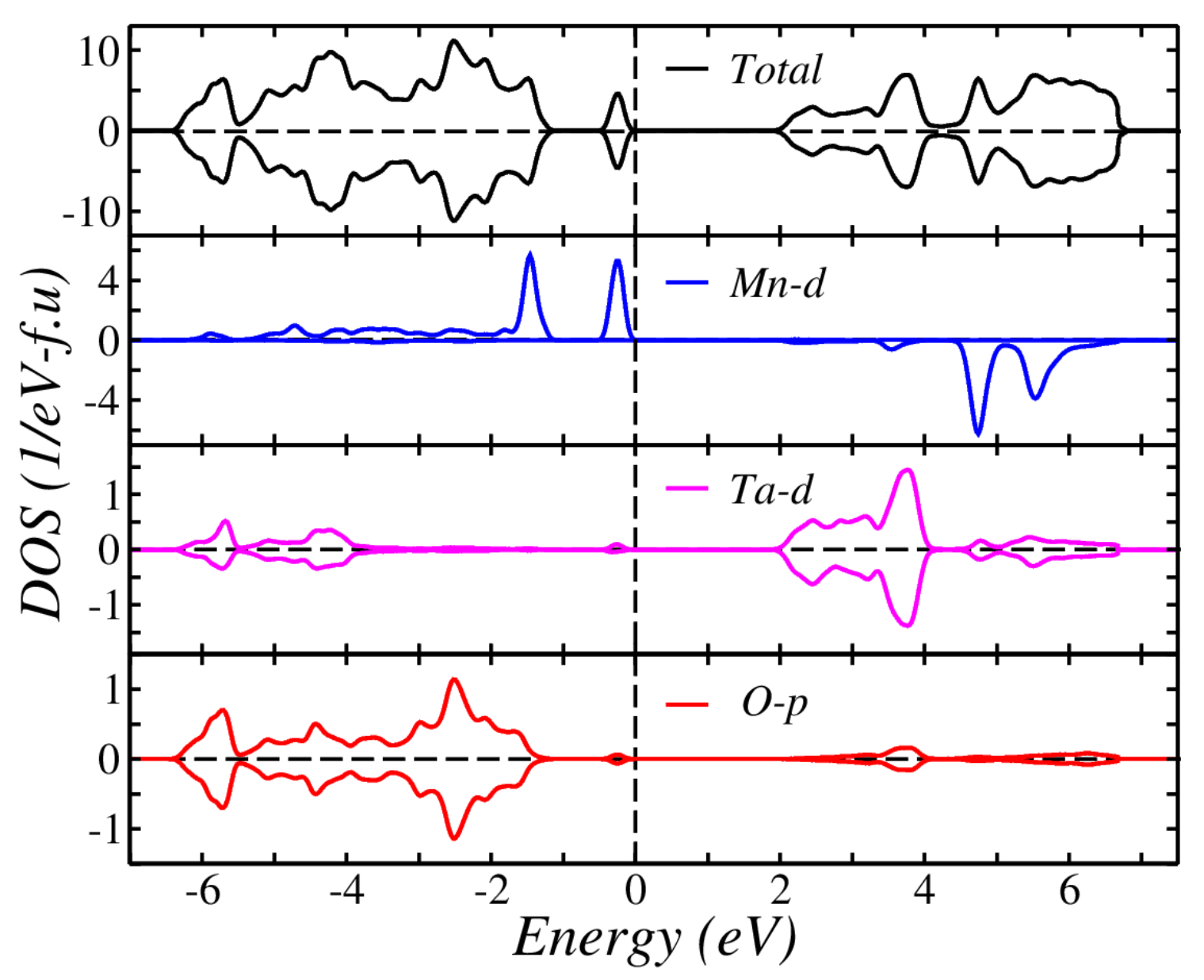}
	\caption{ Spin-polarised total and partial density of states (DOS) in the antiferromagnetic lowest energy state. Here, the Fermi energy is set to 0.0 eV.}
	\label{fig: dos}
		\end{center}
\end{figure}

\begin{table}[h]
	\begin{center}
		\caption{The details of exchange coupling strength as obtained using  GGA+$U$ calculations. The number of nearest neighbor (NN) Mn$^{2+}$ ions corresponding to each interaction is also provided. }
		\vspace{0.2cm}
		\begin{tabular}{ c     c     c     c  }
			\hline
			\hline
			Interaction & \hspace{0.1cm} Bond distance (\AA ) \hspace{0.1cm} & \hspace{0.01cm} Exchange energy (meV) \hspace{0.01cm}  \\
			(NN) &  &   \\
			\hline
			$J_1$ (6) & 5.82 & -1.05 \\
			$J_2$ (2) & 7.13 & -0.02 \\
			$J_3$ (12) & 9.20 & -0.07 \\
			\hline
		\end{tabular}
		\label{tab: exchange}
	\end{center}
\end{table}

\par 
The calculated total density of states (DOS) for the AFM phase (Fig.~\ref{fig: dos}) confirms that BMTO is an insulator with an electronic band gap of 1.9\,eV. The partial density of states (PDOS) analysis reveals that the majority-spin Mn-$d$ states are fully occupied and located well below the Fermi level, while the minority-spin Mn-$d$ states remain entirely unoccupied. This electronic configuration is characteristic of Mn$^{2+}$ in a high-spin ($S$ = 5/2) state, as expected from Hund’s rule. Such a high-spin state is consistent with the effective moment obtained from the Curie-Weiss analysis of our magnetic susceptibility data. The PDOS of the Ta-$d$ states show that they primarily appear above the Fermi energy, indicating that they do not contribute to the magnetism of BMTO. Additionally, the O-$p$ states exhibit a small contribution just below the Fermi level, suggesting weak hybridization with Mn-$d$ states. This implies a minimal itinerancy of the Mn electrons and validates the localized spin model to describe the magnetic interactions of BMTO. This behavior is also observed in other frustrated Mn oxides ~\cite{Shu2023}, where the electronic structure strongly favors localized moments over itinerant magnetism.

\begin{figure}        
	\begin{center}
		\includegraphics[width=1.0\linewidth]{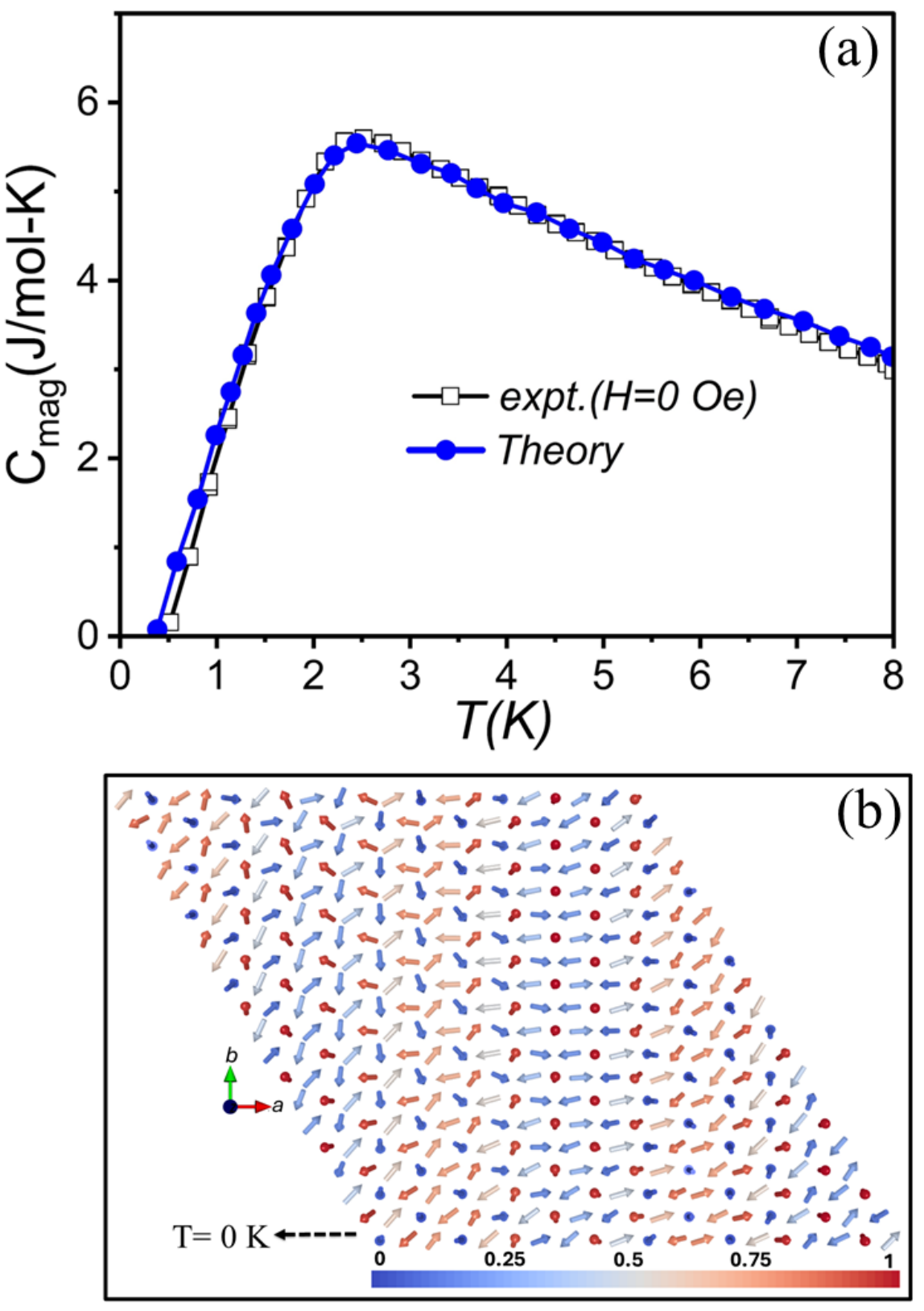} 
		\caption{(a) The magnetic contribution to the specific heat as a function of temperature, $C_{mag}(T)$, calculated from atomistic spin dynamics simulations. (b) The visualization of dynamical ground state spin texture at zero temperature. The color bars associated with the spin textures indicate the projection of the magnetization along the z-direction.}
		\label{fig: CmagDFT}
	\end{center}
\end{figure}

\par 
To gain deeper insight into the magnetic interactions in BMTO, we computed the interatomic exchange parameters $J_{ij}$ as illustrated in Fig. \ref{fig: structure}(b) and summarized in Table \ref{tab: exchange}. Our calculations reveal that the dominant nearest-neighbor (NN) exchange interaction ($J_1$) is strongly antiferromagnetic, consistent with the negative $\theta_{CW}$ as obtained experimentally. This is mediated through a Mn–O–Ta–O–Mn super-superexchange pathway with a Mn–Mn separation of 5.82\,\AA. The second ($J_2$) and third ($J_3$) NN interactions, which couple Mn moments across adjacent layers via a more extended Mn–O–Ta–O–Ta–O–Mn exchange pathway, are considerably weaker with bond distances of 7.13 \AA,  and 9.20 \AA, respectively. The dominance of a strong antiferromagnetic $J_1$ within a triangular geometry naturally leads to frustration, which prevents magnetic ordering even at low temperatures. The hierarchy of exchange interactions suggests that the in-plane frustrated interaction significantly outweighs all other couplings, making it the primary driver of the system’s magnetism (see Fig. \ref{fig: structure}(b) and Table \ref{tab: exchange}). It's worth mentioning that inelastic neutron scattering experiments find the intralayer nearest-neighbor exchange interaction to be antiferromagnetic and 20 times stronger than the interlayer exchange interaction in Ba$_3$MnSb$_2$O$_9$~\cite{Shu2023}. 
\par Our results suggest that the magnetism in BMTO is effectively confined within the two-dimensional (2D) $ab$-plane, making it a part of the intriguing class of quasi-2D frustrated spin systems. The fact that no long-range order is observed in the experiment down to the lowest temperatures measured is in good agreement with the computed nearest neighbor Mn-Mn magnetic coupling which demonstrates a highly frustrated and low-dimensional nature of this material. 
\par 
In order to provide further credence to our estimated magnetic couplings and to understand the temperature-dependent magnetic behaviour, we performed atomistic spin dynamics (ASD) simulations by solving the spin Hamiltonian(\ref{H}) via Stochastic Landau-Lifshitz-Gilbert (SLLG) approach as implemented in UppASD~\cite{PhysRevB.54.1019,Skubic_2008} code. The computed temperature dependent magnetic specific heat ($C_{mag}(T)$) is displayed in Fig.~\ref{fig: CmagDFT}(a). The broad peak in $C_{mag}(T)$ indicates the presence of short-range spin correlations and it agrees well with the experimental data, validating the accuracy of our calculated exchange parameters used in constructing the spin Hamiltonian. This method successfully used to estimate transition temperature for other transition metal compounds~\cite{PhysRevB.111.195148, mallick2025neel}.  The absence of any sharp anomaly in $C_{mag}(T)$ suggests the lack of conventional long-range magnetic order, a characteristic feature of geometrically frustrated spin systems. This behavior, coupled with the dominant in-plane exchange interactions and weak interlayer couplings, underscores the effective two-dimensional (2D) nature of the system. The spin-texture at $T \thickapprox 0$ is calculated and the orientation of Mn$^{2+}$ spin moments in the $ab$ plane, as shown in Fig.~\ref{fig: CmagDFT}(b), further confirms the absence of any long-range ordering in this system. The absence of any long-range ordering could be attributed to the reduced dimensionality and the strong magnetic frustration in this system. Similar frustration-driven physics has been observed in a variety of triangular and honeycomb-based systems~\cite{PhysRevB.81.214419,PhysRevB.84.094424}, where competing interactions often stabilize unconventional magnetic phases, such as spin spirals, spin nematics, or spin liquid states. Given the negligible contribution of spin-orbit coupling (SOC) and the absence of Dzyaloshinskii–Moriya (DM) interactions in BMTO, the system is well-described by a purely isotropic Heisenberg Hamiltonian. The presence of strong in-plane frustration, and quasi 2D nature suggests that the present system could be a promising candidate for realizing nontrivial magnetic ground states, warranting further theoretical and experimental analysis of its low temperature behavior. 

\section{Conclusion}
We have synthesized and studied in detail the magnetic properties of a new triple perovskite material Ba$_3$MnTa$_2$O$_9$. The material crystallizes in the trigonal structure ($P\overline{3}m1$) with a high degree of cation order. The only magnetic ions (Mn$^{2+}$) are in a high-spin state ($S$ = 5/2) and form a quasi-2D triangular lattice parallel to the crystallographic $ab$ plane. A Curie-Weiss temperature ($\theta_{CW}$ = -12\,K) obtained from the fitting of susceptibility data suggests the spins are antiferromagnetically coupled. This is also confirmed by our DFT+U calculations. Low-temperature susceptibility and specific heat data show that the system does not order down to at least 0.5\,K, and the absence of spin glass behavior is confirmed by ac-susceptibility measurements. Our theoretical calculations provide microscopic insight into the experimentally observed absence of magnetic ordering in BMTO. By computing interatomic magnetic interactions and performing spin dynamics simulations, we have established a consistent picture of frustration-induced suppression of ordering. A broad maximum found at $\sim$2\,K in the magnetic specific heat ($C_{mag}$) and magnetic susceptibility, as well as an increase in the ESR linewidth with decreasing temperature, suggest the presence of short-range spin correlations. The entropy of the material is released over an extended temperature range, further indicating its frustrated nature. All these results suggest that BMTO could potentially host a novel and exotic magnetic state, driven by frustration and reduced dimensionality. These findings warrant further experimental and theoretical investigations to understand the spin excitations and ground-state properties of the material. 

\section*{ACKNOWLEDGMENTS}
TD sincerely acknowledges the financial support provided by IIT (ISM) Dhanbad through the Faculty Research Scheme (FRS(171)/2022-2023/PHYSICS), and Anusandhan National Research Foundation (ANRF), India through a start-up research grant (SRG/2022/001339). RM and TD express their gratitude to DST India and DAAD Germany for facilitating their visit to IFW Dresden, Germany, through the Indo-German DST-DAAD personnel exchange program (DST/ICD/GERMANY/DAAD/2023/131). SN acknowledges the funding support for Chanakya Postdoctoral Fellowship (CPDF/2021-22/01) from the National Mission on Interdisciplinary Cyber Physical Systems of the DST, Govt. of India through the I-HUB Quantum Technology Foundation. SKP acknowledges support from ANRF (previously
SERB), Government of India for the core research grant (CRG/2023/003063). The work in Dresden was supported by the Deutsche Forschungsgemeinschaft (DFG, German Research Foundation) through Grant No. 455319354, Grant No. WU595/14-1, Project No. 247310070 (SFB 1143), Project No. 247310070 (SFB 1143), and Project No. 390858490 (W\"urzburg-Dresden Cluster of Excellence \textit{ct.qmat}, EXC~2147).

%
%
%
%
%
%
%
%
%
%
%
%
%
%
%
%
%
%
%
%

\bibliography{ref}

\end{document}


\title{Supplementary information: Frustration-driven unconventional magnetism in the Mn$^{2+}$ ($S=\frac{5}{2}$) based two-dimensional triangular-lattice antiferromagnet Ba$_{3}$MnTa$_{2}$O$_{9}$}

\author{Romario Mondal}

\affiliation{Department of Physics, IIT (ISM) Dhanbad, Jharkhand 826004, India}

\author{Sk. Soyeb Ali}

\affiliation{Department of Physics, Bennett University, Greater Noida, Uttar Pradesh 201310, India}

\author{Saikat Nandi}

\affiliation{Department of Physics, IIT Bombay, Powai, Mumbai 400076, India}

\author{S. Chattopadhyay}

\affiliation{UGC-DAE Consortium for Scientific Research Mumbai Centre, 246-C CFB, BARC Campus, Mumbai 400085 India}

\author{S. Ga\ss}

\affiliation{Leibniz Institute for Solid State and Materials Research Dresden, Dresden D-01069, Germany}

\author{L. T. Corredor}

\altaffiliation{Present address: Faculty of Physics, Technical University of Dortmund, Otto-Hahn-Str. 4, D-44227 Dortmund, Germany, and Research Center Future Energy Materials and Systems (RC FEMS), Germany}

\affiliation{Leibniz Institute for Solid State and Materials Research Dresden, Dresden D-01069, Germany}

\author{A. U. B. Wolter}

\affiliation{Leibniz Institute for Solid State and Materials Research Dresden, Dresden D-01069, Germany}

\author{V. Kataev}

\affiliation{Leibniz Institute for Solid State and Materials Research Dresden, Dresden D-01069, Germany}

\author{B. B\"uchner}

\affiliation{Leibniz Institute for Solid State and Materials Research Dresden, Dresden D-01069, Germany}

\affiliation{Institute for Solid State and Materials Physics and W{\"u}rzburg-Dresden Cluster of Excellence ct.qmat, TU Dresden, D-01062 Dresden, Germany}

\author{A. Alfonsov}

\affiliation{Leibniz Institute for Solid State and Materials Research Dresden, Dresden D-01069, Germany}

\author{S. Wurmehl}

\affiliation{Leibniz Institute for Solid State and Materials Research Dresden, Dresden D-01069, Germany}

\author{A. V. Mahajan}

\affiliation{Department of Physics, IIT Bombay, Powai, Mumbai 400076, India}

\author{S. K. Panda}
\email[Email: ]{swarup.panda@bennett.edu.in}
\affiliation{Department of Physics, Bennett University, Greater Noida, Uttar Pradesh 201310, India}

\author{T. Dey}
\email[Email: ]{tushar@iitism.ac.in}

\affiliation{Department of Physics, IIT (ISM) Dhanbad, Jharkhand 826004, India}

\maketitle

\subsection*{Analysis of crystal structure}
We used the following datasets from the Inorganic Crystal Structure Database (ICSD) for comparison: Ba$_3$MnNb$_2$O$_9$ in the cubic (ICSD no 230384; space group \#225, $Fm\overline{3}m$) and trigonal variant (ICSD no 230383; space group \#164, $P\overline{3}m1$), as well as Ba$_5$Ta$_4$O$_{15}$  (ICSD no 16028; space group \#164).
\renewcommand{\thefigure}{S\arabic{figure}}  
\setcounter{figure}{0}                       
\begin{figure}[h]
	\begin{center}
	\includegraphics[width=\linewidth, trim={0cm 8cm 8.5cm 0cm,clip}]{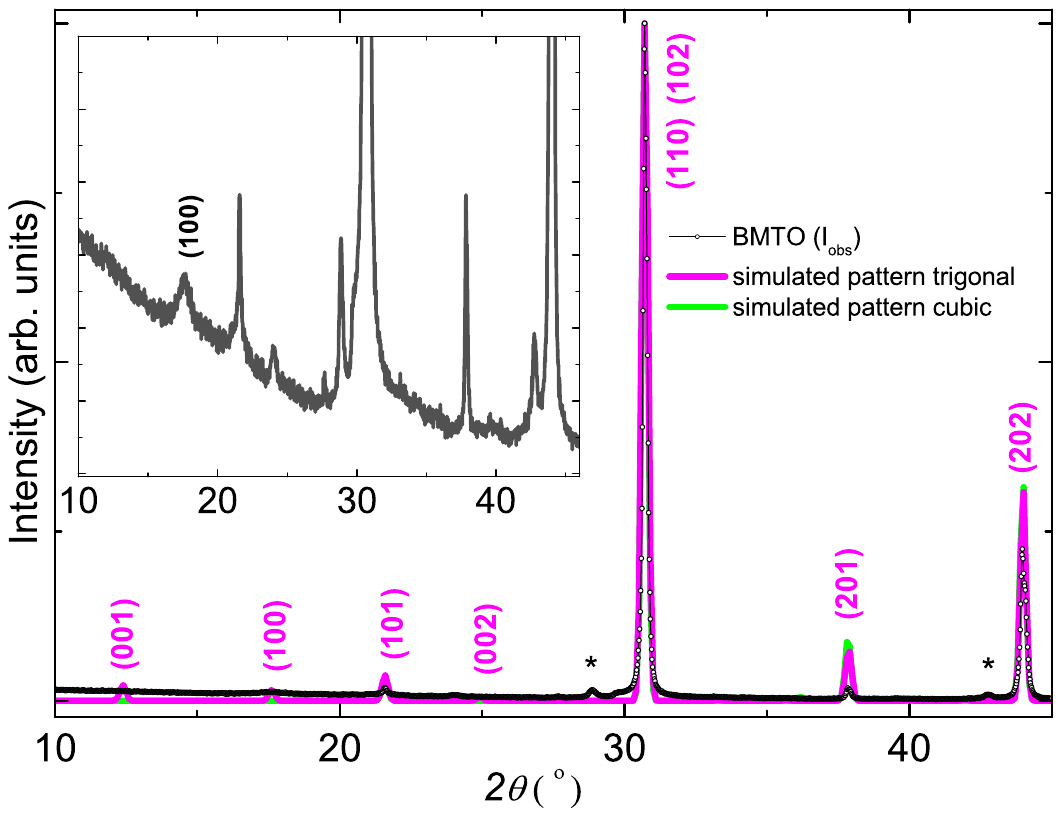}
	\caption{The simulated trigonal (magenta) and cubic structure-type (green) XRD patterns of BMTO in comparison with the experimental data. Please note that all intensities were normalized with respect to the peak height of the main (110/102) reflection at 30.9$^\circ$. The inset shows a zoomed low-angle region and indicates the broadened (100) reflection.}
	\label{fig:S1}
	\end{center}
\end{figure}

\begin{figure}[h]
	\begin{center}
	\includegraphics[width=\linewidth, trim={1cm 8cm 9cm 0cm,clip}]{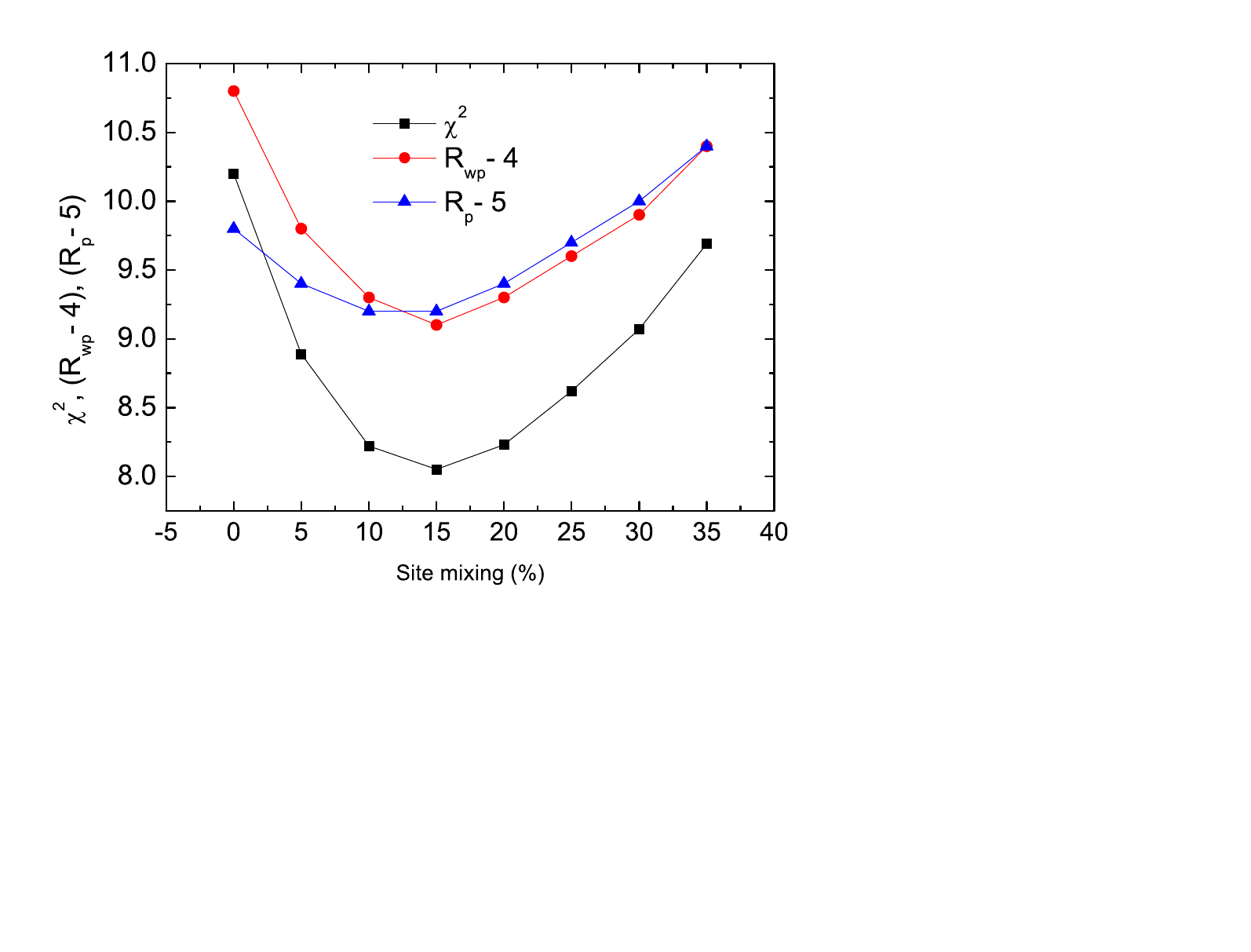}
	\caption{The variation of $\chi^2$, $R_{wp}$, and $R_p$ with the anti-site mixing (\%) between Mn$^{2+}$ and Ta$^{5+}$ is shown. $R_{wp}$ and $R_p$ values are offset down by 4 and 5, respectively, for clarity.}   
	\label{fig:S2}
	\end{center}
\end{figure}

Two possible structure polymorphs have been reported in the case of BMTO. (i) The fully B-site ordered variant is described by the formula Ba$_3$MnTa$_2$O$_9$ (trigonal symmetry, space group \#164), where Mn and Ta are ordered in a way that the sequence of B-site atoms is ...Mn-Ta-Ta-Mn-Ta-Ta... Here Mn occupies the Wyckoff position 1$b$ only, whereas Ta solely fills the Wyckoff position 2$d$, consistent with the overall 1:2 stoichiometry of Mn:Ta. (ii) The partially disordered variant (Ba$_3$MnTa$_2$O$_9$)$_{1/3}$, where Ta fully occupies the Wyckoff position 4$b$ whereas  $2/3$ and $1/3$ of the Wyckoff position 4$a$ is filled with Mn and Ta, respectively,  which are randomly distributed (cubic symmetry, space group \#225). Both variants are differentiated by the presence or absence of several reflections, albeit their difference is rather subtle as outlined, e.g., in Refs. \cite{lee2014, xin2018} (also compared in Fig. \ref{fig:S1}).
\par
Our synthesis conditions suggest predominant formation of the ordered variant in line with the structure analysis by Treiber \textit{et al.} who reported the formation of the ordered, trigonal Ba$_3$MnTa$_2$O$_9$ above 1200$^{\circ}$C \cite{TK82}. However, as our synthesis protocol uses slow cooling, this may, in part, lead to the formation of the low-temperature disordered phase \cite{TK82}. Our rationalization is also consistent with the structure analysis of the sister compound Ba$_3$MnNb$_2$O$_9$, which has been reported to crystallize in an ordered trigonal crystal system with space group $P\overline{3}m1$ \cite{xin2018}. Given that the ionic radii of Nb$^{5+}$ and Ta$^{5+}$ in an octahedral environment ($r$ = 0.64\,\r A \cite{Shannon1976}) are very close, a structure model similar to Ba$_3$MnNb$_2$O$_9$ was used as an educated guess to fit the room-temperature XRD pattern of BMTO.
\par
Any change in crystal symmetry might affect the exchange paths. Hence, to eliminate any doubt about the symmetry of the lattice, we carefully verified the possibility of a cubic variant as well. We simulated powder patterns using the reported structure data of Ba$_3$MnNb$_2$O$_9$ for the cubic (ICSD no 230384; space group \#225, $Fm\overline{3}m$) and trigonal variants (ICSD no 230383; space group \#164, $P\overline{3}m1$) and compared with our experimental data on BMTO (see Fig. \ref{fig:S1}). We have chosen the data of the Nb-based sister compound as a reference since both the cubic and trigonal phases have been reported in one publication, suggesting a consistent analysis \cite{xin2018}. 
It has been earlier reported using a combination of x-ray diffraction, neutron diffraction, and electron microscopy experiments that there is a correlation between composition, domain size, and the extent of cation order within the respective domains in Ba$_3$ZnTa$_2$O$_{9-\delta}$ even including formation of two trigonal Ba$_3$ZnTa$_2$O$_{9-\delta}$ phases with different degrees of cation ordering within one sample \cite{BMN03}. The most remarkable result in reference \cite{BMN03} is that the sample with the highest degree of cation ordering within the trigonal lattice is the one with a broad (100) reflection. 
At first glance, a comparison between the XRD pattern of the Ba$_3$MnNb$_2$O$_9$ model systems in trigonal and cubic symmetry and the experimental data of BMTO (see Fig. \ref{fig:S1}) reveals that the (001) reflection is absent, whereas the (100) reflection is clearly observed (see inset of Fig. \ref{fig:S1}) albeit reduced in intensity and broadened significantly as compared to the other reflections. Following the analysis of Bieringer \textit{et al.} \cite{BMN03}, the presence of the (100) superstructure reflection (2$\theta$ $\approx$ 17.7$^{\circ}$) as such indicates that most of the Mn$^{2+}$ and Ta$^{5+}$ ions are ordered whereas the significantly broadened width originates from a small size of the structure domains of the trigonal perovskite phase \cite{BMN03}. 
\par
For further insight into the broadening of superlattice reflections, especially the (100) reflection, the following procedure was adopted. After the double-phase refinement of the XRD data, the occupancies of Mn$^{2+}$ and Ta$^{5+}$ in the sites $1b$ and $2d$, respectively, were varied by 5\% in each step and refined carefully by maintaining the general triple-perovskite formula of BMTO. Please note that at every step, all other parameters were kept fixed. As shown in Fig. \ref{fig:S2}, a minimum appears at $\sim$15\% in the goodness of fitting parameters $\chi^2, R_{wp}-4$, and $R_p-5$ vs. \% site mixing graph. Here, $\chi^2$, $R_{wp}$, and $R_{p}$ are the weighted squared difference between the observed and calculated pattern, the weighted profile factor, and the profile factor, respectively. This result suggests an anti-site mixing of $\sim$15\%. This is another indication of the possible superlattice peak broadening. The values of site occupancies corresponding to this minimum are provided in Table I of the main paper. Based on the arguments given above, we conclude that the BMTO sample under study orders in the trigonal structure, and has a high degree of cation order. Using the preconditions, we are able to refine a structure model that describes our BMTO data at low angles reasonably well.


\bibliography{ref}